\documentclass[aps,reprint,onecolumn,notitlepage,superscriptaddress]{revtex4-1}
\usepackage{graphicx}
\usepackage{times}
\usepackage[colorlinks=true, allcolors=blue]{hyperref}
\makeatletter
\bibpunct{}{}{,}{s}{}{,}
\begin{document}
%\newcommand{\hdblarrow}{H\makebox[0.9ex][l]{$\downdownarrows$}-}
%\linenumbers
\title{Complex beam mapping and fourier optics analysis of a wide field Microwave Kinetic Inductance Detector camera}
\date{\today}

\author{S. J. C. Yates}
\email[e-mail: ]{s.yates@sron.nl}
\affiliation{SRON Netherlands Institute for Space Research, Landleven 12, 9747 AD Groningen, The Netherlands}
\author{K. K. Davis}
\affiliation{University of California Santa Barbara, Santa Barbara, CA, USA}
\author{W. Jellema}
\affiliation{SRON Netherlands Institute for Space Research, Landleven 12, 9747 AD Groningen, The Netherlands}
\affiliation{Kapteyn Institute, University of Groningen, Groningen, 9747 AD The Netherlands}
\author{J. J. A.  Baselmans}
\affiliation{SRON Netherlands Institute for Space Research, Utrecht, 3584CA, The Netherlands}
\affiliation{Terahertz Sensing Group, Delft University of Technology, Delft 2628CD, The Netherlands}
\author{A. M. Baryshev}
\affiliation{Kapteyn Institute, University of Groningen, Groningen, 9747 AD The Netherlands}

\begin{abstract}
For astronomical instruments, accurate knowledge of the optical pointing and coupling are essential to characterize the alignment and performance of (sub-)systems prior to integration and deployment. Ideally, this requires the phase response of the optical system, which for direct (phase insensitive) detectors was not previously accessible. Here we show development of the phase sensitive complex beam pattern technique using a dual optical source heterodyne technique for a large field of view Microwave Kinetic Inductance Detector camera at 350~GHz. We show here how you can analyze the measured data with Fourier optics, which allows integration into a telescope model to calculate the on sky beam pattern and telescope aperture efficiency prior to deployment at a telescope. 
\end{abstract}

\maketitle
\section{Introduction}
Traditionally near field optical beam pattern measurements with thermal (incoherent) sources~\mbox{\cite{Murphy2010-multimodePlanckshort}} have been used to characterize phase-insensitive broadband (direct)  detectors. However, complex (coherent) field mapping of both amplitude and phase patterns offers several advantages~\cite{Teyssier2008-HIFIpre-flight-testing,Tong2003-NFbeams1THzshort,Tervo2002-APWS-coherentEMfields} in terms of higher signal to noise and allowing Fourier optics analysis using the measured beam patterns~\cite{IEEEMTT83Martin}. While the complex beam pattern~\cite{Tong2003-NFbeams1THzshort} technique is not typically used for direct detectors it is the standard characterization tool for missions using heterodyne receivers, and has been used to characterize instrumentation from small, ground-based missions to the large and satellite-class missions such as IRAM~\cite{Carter:ISSTT02short}, Herschel HIFI~\cite{Jellema-PhD2015}, ALMA~\cite{Naruse2009-ALMAb8NFbeamsshort,ALMAband9short}. 

The approach followed here with microwave kinetic inductance detectors~\cite{PeterK.Day2003-NatureMKIDsshort} (MKIDs) was first presented by Davis et al. 2017~\cite{Davis2017-ConceptComplexMKIDsshort} using a heterodyne technique. This approach was expanded upon in  Davis et al. 2019~\cite{Davis2018-LargeComplexMKIDsshort}, combining the technique with wide field optics, multiplexed electronics and a large array of MKIDs. This paper expands on that work, showing how a complex beam pattern from an instrument can be used prior to integration and deployment to calculate the on sky beam pattern and coupling efficiencies with Fourier optics~\cite{IEEEMTT83Martin}, similar to that done for Herschel HIFI~\cite{Jellema-PhD2015}.  

\section{Experimental methods}
\begin{figure*}
  %\centering
  \includegraphics[width=\textwidth]{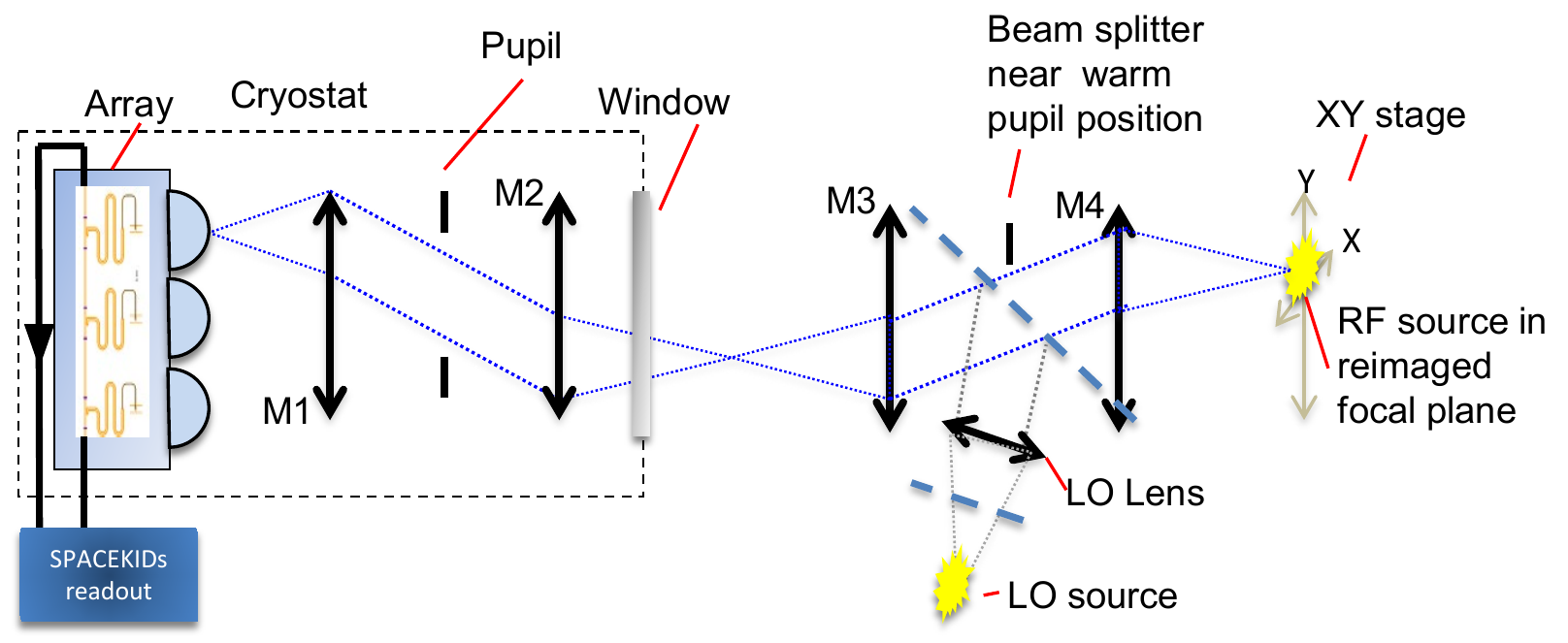}
  \caption[Scheme for FPA Complex Field Measurements]
  {(color online) Paraxial approximation of optical coupling scheme. The dotted blue line shows the beam of one pixel, the black lines and arrows show active optical components. We measure the optical beating of two signals, from the local oscillators (LOs), one which is fixed and illuminates the entire array ($LO$) and one which scanned ($RF$). A phase reference is generated separately by mixing the signal generators together, which is acquired by mixing with the MKID readout.}
\label{fig:schema}
\end{figure*} 
In this paper we only summarize the experimental setup for this measurement, but it is covered in detail in Davis et al.~\cite{Davis2018-LargeComplexMKIDsshort}. As an example for the analysis we use data from a lab based reflective optics based large field camera with a base T of 240~mK, see Ferrari et al.~\cite{IEEE18Ferrari_testbedshort} for more details. An overview of the optics is shown in Fig.~\ref{fig:schema}: it can be considered as two Gaussian beam telescopes~\cite{GoldsmithPaulF1998Qs:G} or relays with a total system magnification of 3. Each relay is made of two active mirrors (M1 to M4) separated by their focal lengths, with an additional three fold mirrors. One relay is placed at 4~K and has an aperture stop the ``pupil" which limits opening angle on the array at 14~deg. or f\#2, f\# the focal length to beam diameter ratio. A second relay is placed in the warm, giving access in the lab to the (aberration compensated~\cite{murphy:IJRMW87}) instrument focal plane.  The field of view is $\sim6\times6$~cm at the array, similar to the array size. The array used has 880 pixels lens-antenna coupled for 350~GHz with a stray light absorber on chip to kill in chip stray light (or the "surface wave")~\cite{Yates2017short}. Each pixel is a hybrid quarter wavelength MKID made of NbTiN with an active portion for radiation detection of 40~nm thick Al. The entire array is on a single readout line from 4--8~GHz. For more details on the design, fabrication and optical verification see Yates et al.~\cite{Yates2017short}. The pixels have a separation of 2~mm, or 1.2~f\#$\lambda$ so the beams are designed to overlap to improve mapping speed~\cite{Griffin2002-ApplOptics-Feedhorn-coupling}. As a consequence the beams over-illuminate the pupil, having a relatively high edge taper of $\sim-5$~dB and consequently a focal plane beam pattern with the first sidelobe at $\sim-18$~dB. 

\begin{figure*}
	\includegraphics[width=\textwidth]{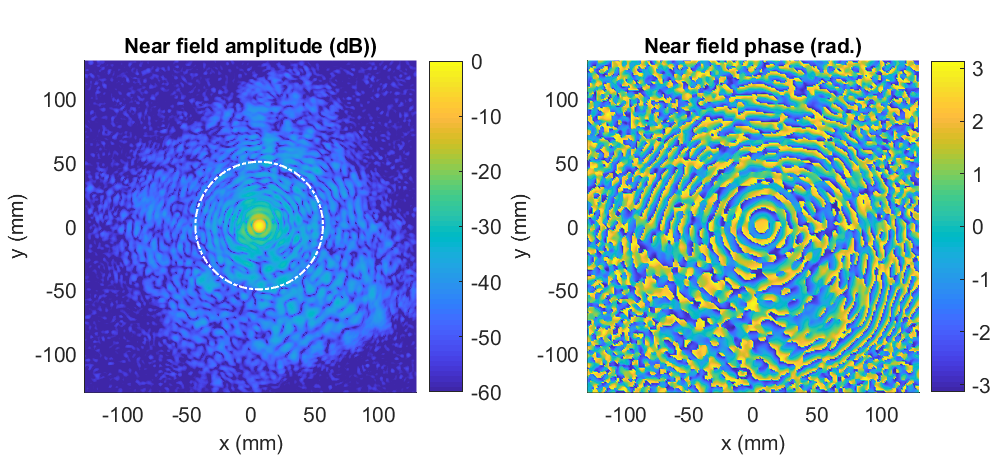}
	\caption[~Near field Data, Amplitude and Phase]
    {(color online) Full processed and spatially filtered complex field map (in dB) for a representative pixel located at the center of the array, spatially filtered and propagated by 30~mm from the measurement plane to the waist position. The data shown is shown for the co-polarization. The rotated square with amplitude signal at $\sim$ -40 dB shows the extent of the field of view before dropping off to the noise floor at $\sim$-60~dB. The dotted circle indicates the focal plane mask to select data prior to further analysis.}
    \label{fig:copol_nf}
\end{figure*}

The technique presented is based on the measurement of the optical interference between two optical sources in a heterodyne configuration (see Fig.~\ref{fig:schema}). One source, the local oscillator "LO", is coupled to the entire array using a thin beamsplitter placed a warm pupil position. The second source, the radio frequency source "RF", is scanned in the focal plane in a similar fashion to previous traditional beam pattern measurements. Both sources are driven at $\times$~32 harmonic of the input synthesisers. As the RF source is scanned, we measure the optical interference~\cite{ThomasWithington2012-MultiModeCoherentTesting} between the detected RF source, which is convolved with the full system beam pattern, and the LO source. The optical interference is moved away from DC to reduce 1/f noise by offsetting the optical sources by a small frequency (IF$_{mod}$=565~Hz), which has to be within the response bandwidth of the detector and readout. A phase reference signal is created by mixing the signals from the synthesizers to produce a signal at $\Delta f=\mathrm{IF}_{mod}$/32=17.6~Hz,  acquired as an extra virtual pixel by mixing it with the readout. A processed example phase and amplitude beam pattern is shown in Fig.~\ref{fig:copol_nf}. Since we use a multiplex readout, $\sim$400 pixels are measured with one readout~\cite{rantwijk:IEEE16short} so the most of the array can be measured in two scans. Two separate scans with the RF source in orthogonal polarization orientations are measured, which are then phase shifted and combined in post-processing to extract the co and cross-polarizations beam patterns (see Davis et al. 2018~\cite{Davis2018-SPIE-cpxTechniquesshort}). The data presented and analyzed in the following sections are the co-polarization from Davis et al. 2019~\cite{Davis2018-LargeComplexMKIDsshort}.

\section{Analysis}
\begin{figure*}
  \centering
  \includegraphics[width=\textwidth]{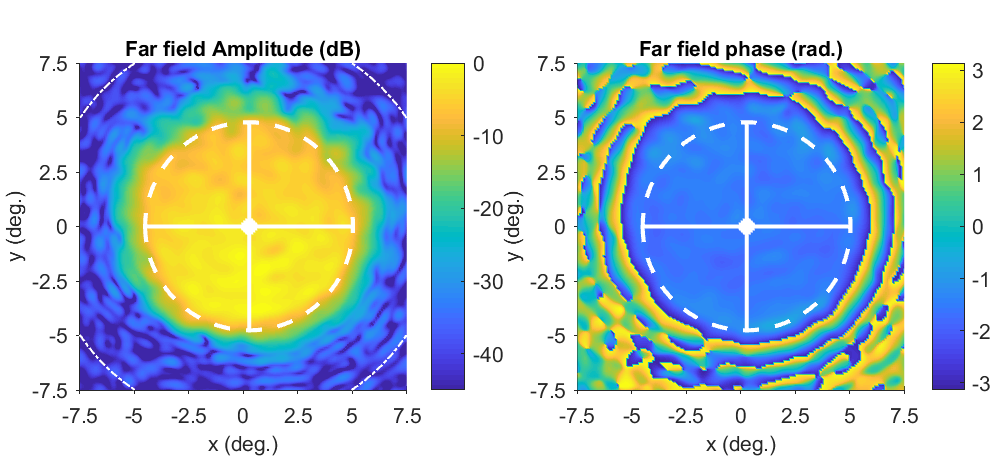}
  \caption[~Far Field Transformation] 
  {(color online) Far-field co-polarization radiation patterns (dB) of a representative MKID pixel, with focal plane mask applied. The white dashed circle indicates the designed optical aperture corresponding to f\#=6, (f\# the focal length over diameter ratio). The mask for telescope secondary and support struts are additionally shown in white. The dotted circle indicates the spatial filter applied to the data prior to further analysis.}
  \label{fig:copol_ff}
\end{figure*}
A first order analysis is to fit the measured data to a Gaussian beam~\cite{GoldsmithPaulF1998Qs:G} which is presented in Davis et al.~\cite{Davis2018-LargeComplexMKIDsshort}. This analysis gives the actual focus position, waist size and Gaussicity which is the coupling to an optimum Gaussian beam. A further analysis is to then calculate the far field beam pattern with a Fast Fourier Transform (FFT) of the data~\cite{IEEEMTT83Martin}, shown in Fig.~\ref{fig:copol_ff}. We expect a truncated Gaussian in this plane, and we see in the amplitude the pupil aperture at the expected angular extent of 4.8~deg. or f\#6. The phase has been flattened to correct for offsets in x, y and z. We can then calculate the wavefront error (WFE) from the rms of the phase over the entire pupil which is 23~$\pm~3~\mu$m, including phase error from the measurement. Since we have a non-uniform illumination, of more interest is the amplitude weighted wavefront error of 10~$\pm~2~\mu$m. Note, this is the wave front error, the combined shape error of the seven mirrors is therefore half this number. The WFE can be expressed as an efficiency from \cite{ruze66}, $\eta_{ruze}=exp(-(2\pi\sigma)^2/\lambda^2)$. Taking $\eta_{ruze}=0.8$ as the lowest acceptable efficiency, this wavefront error would imply the optics work with a similar illumination up to a frequency of $\sim$2.3~THz, far surpassing the design goal of 900~GHz assuming the errors have a larger length scale than the wavelength.

The noise in the near field beam can be reduced by applying a ``spatial filter", a mask on the far field which selects the beam only up to a certain angle. The spatially filtered near field beam pattern is then just the inverse FFT of the spatially filtered far field beam. The data in Fig.~\ref{fig:copol_nf} has such a filter applied at 9~deg. which reduces the noise floor by about 5~dB to the -60~dB level. Further, by applying a properly scaled phase transformation to the far field the beam can be propagated to any required position in the optics~\cite{IEEEMTT83Martin}. We use this in Fig.~\ref{fig:copol_nf} to propagate 30~mm from the measurement plane to the focal plane (the waist position). 

The LO coupling is not fully uniform across the array, changing the modulation depth across the field of view and hence noise floor from $-55$~dB to $-30$~dB relative to the beam maximum. Additionally, the array is optimized for lower background ($\sim50$~K) than under the LO illumination ($\gg300$~K) so some pixels ($\sim5\%$) have high readout noise contribution. So to investigate only the optical coupling the data in the following sections has both the spatial filter (at 9~deg. half width) and a focal plane mask at 50mm radius to limit noise contribution. A cross check with pixels with good signal to noise shows this underestimates the power in the residual stray light by $\sim2\%$, visible as a signal at $\sim-40$~dB in Fig.~\ref{fig:copol_nf} in the field of view but off pixel.

\section{Telescope coupling}  
\begin{figure}
  \centering 
	\includegraphics[width=\textwidth]{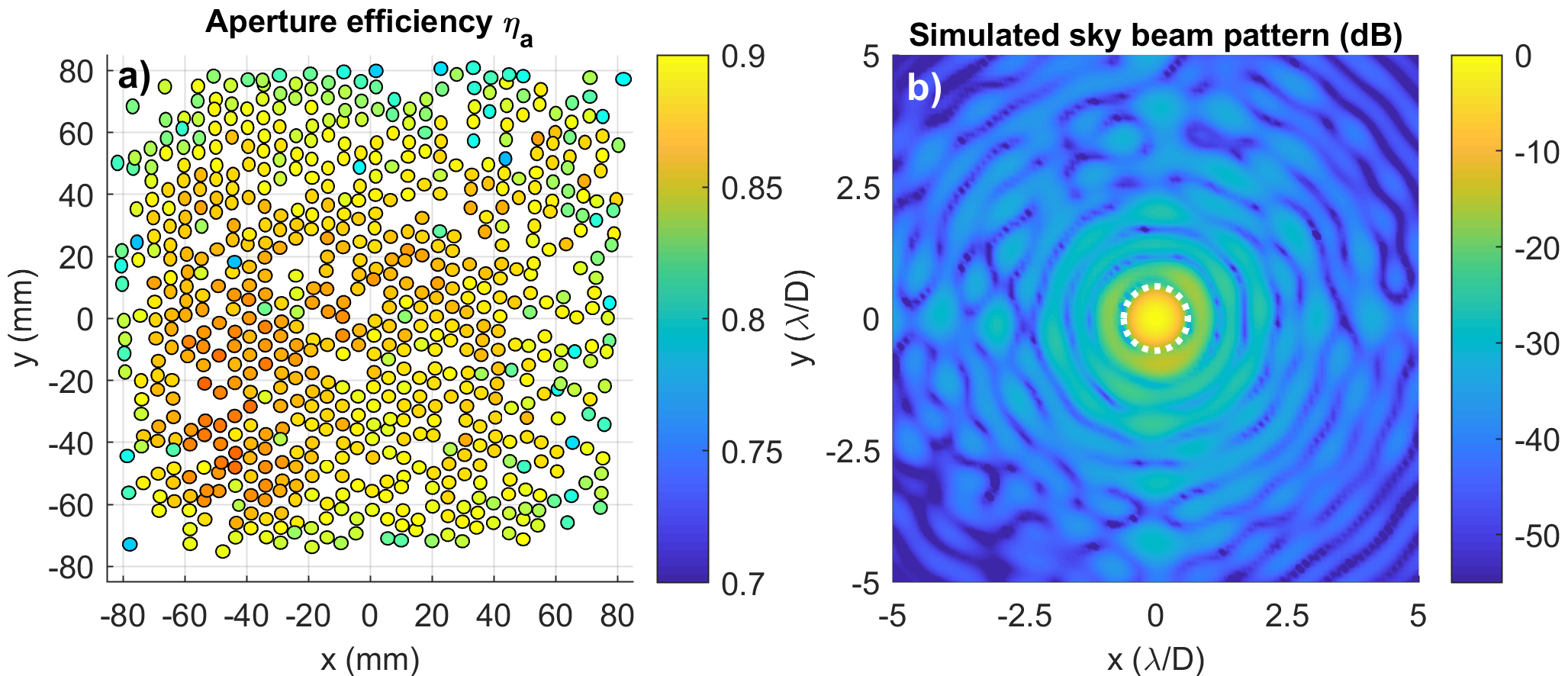}
	\caption[~Coupling efficiencies]
    {(color online) Left a): aperture efficiency as a function of position. Right b): Calculated on sky beam pattern, calculated using near field laboratory complex field beam pattern measurements. The white circle indicates the diffraction limit of $1.22 \lambda/D$, D the telescope diameter. See text for more details.}
    \label{fig:cfeta}
\end{figure}

As an exercise we can therefore see how our lab-based instrument would couple to a telescope using Fourier optics with the measured beam patterns. Here, we take as a model that the focal plane of the instrument is at the focal plane of the telescope, typically a Cassegrain system made of a small secondary and large primary mirror. As a further simplification, we can take the measured far field beam pattern as the near field pattern of the telescope. We can then apply a mask, shown on the far field amplitude in Fig.~\ref{fig:copol_ff}, to simulate the finite size of the secondary (taken at f\#6, so matching the designed system), the blockage due to the secondary mirror and support struts. The field curvature of this instrument is similar to one that might be integrated in a telescope, so is fitted and removed giving the angular offset seen in Fig.~\ref{fig:copol_ff}. Alternatively the field curvature can be constrained to simulations or by propagating the beam to the secondary position. The on sky beam pattern is then the FFT of the masked illumination, where the angular scale is given by the $\lambda/D$ with D the primary dish size, see Fig.~\ref{fig:cfeta}b).

We now consider several efficiencies defined as in~\cite{GoldsmithPaulF1998Qs:G}: the spillover efficiency  $\eta_f$, the fraction of the beam incident on the antenna aperture, here we also include the blockage efficiency; the taper efficiency $\eta_{t}$, the fractional coupling over the antenna area ignoring spillover, this includes phase losses and the illumination; the aperture efficiency $\eta_a=\eta_f\times\eta_{t}$, the coupling to a incident plane wave. These efficiencies are referenced at the primary mirror for an incident plane wave, so describe coupling to a point source. Alternatively we can relate the efficiencies to the on sky beam pattern which is useful for finite size sources. For this we must consider the filling fraction of the source to the beam, the main beam pattern efficiency  $\eta_{mbp}$. We can further expand on this with the main beam efficiency, where we include also the spillover on the secondary $\eta_{mb}=\eta_{mbp}\times\eta_f$. For reference we take here the main beam efficiency for a source of radius of $1.22\lambda/D$, so excluding the sidelobes (see Fig~\ref{fig:cfeta}b), this is similar to the aperture efficiency for a point source as might be expected. The efficiencies are over-viewed in Tab.~\ref{tab:eff} and Fig.~\ref{fig:cfeta} for the entire array and compared to simulations using physical optics. We see the performance gives close to the theoretical (near aberration free) with little position dependence (see Fig.~\ref{fig:cfeta}). Yield is limited by the accuracy of the MKID frequency placement and the resulting crosstalk from overlapping MKID dips (see for example Baselmans et al.~\cite{baselmans:AA17short} and references there in), but still near the target 80\%. There is a small loss of main beam efficiency due to increased spillover (0.87 versus the designed 1) on the secondary, which is seen in Fig.~\ref{fig:copol_ff} as some signal outside the pupil aperture. Analysis shows this to be a pupil aperture defocus which was introduced in the design due to the difficulty in folding the optics. However, this spillover would be on the sky and not to 300~K so this is only a loss in efficiency and not an increase in stray loading which would give a larger sensitivity problem. Improved baffling around the pupil would mitigate this problem. Note the efficiency quoted is for the measured beam pattern, losses between this position and the array are not included. However, you can take the sensitivity of the system at the same position as where the beam pattern is measured and then combine it with the above efficiencies to get the full system point source (or extended source) sensitivity prior to integration and deployment. Note that a real telescope will also have a finite surface roughness which could be separately measured and added to this model, here we take an ideal telescope. 

\begin{table}
\caption{Optical efficiency on simulated sky for measured beam pattern, see text for details. These are for the measured beam patterns, so in addition to cold losses such as the lens antenna efficiency and cold optics. The main beam $\eta_{mb}$ and beam pattern $\eta_{mbp}$ efficiencies are calculated for a source filling the main beam, so a radius of ${1.22\lambda/D}$. The range given is the standard deviation of the $\sim90\%$ best pixels of those measured array (667 of 718 measured).} 
\label{tab:eff}
  \centering 
\begin{tabular}{c | c c}        % centered columns (4 columns)
\hline \hline
Efficiency & Simulated & Measured \\
\hline \hline 
Gaussicity & 0.91 & 0.87 $\pm$ 0.01 \\
Aperture  $\eta_a$ & 0.9 & 0.82 $\pm$ 0.02  \\
Taper  $\eta_{t}$ & 0.9 & 0.92 $\pm$ 0.02\\
Spillover  $\eta_f$ & 1 & 0.90 $\pm$ 0.02 \\
Main beam pattern  $\eta_{mbp} $  & 0.89  & 0.9 $\pm$ 0.01\\
Main beam  $\eta_{mb}$ & 0.89 & 0.80 $\pm$ 0.02\\
\end{tabular}
\end{table}

\section{Conclusions}
We present here complex beam measurements on a large field MKID camera at 350~GHz. We showed how this can be used to gain understanding of the system which would not be possible by any previous technique used for direct detectors, here identifying spillover from a small pupil aperture shift. This technique opens up the possibility that future direct detector instruments can be characterized to the quality previously implemented as for heterodyne spectrometer missions, such as Herschel HIFI (see~\cite{Jellema-PhD2015}). This paper overviews the techniques in detail used recently by some of the authors to explain the on sky beam pattern and coupling efficiency for the MKID on-chip spectrometer DESHIMA~\cite{NatAst19:Endoshort}. For analyzing more complicated optical systems, the measured complex beam patterns could be alternatively be fully propagated, for example in commercial optics software. Finally, the measurements here were taken near the focal plane, but in principle can be taken at any plane that is accessible and used to calculate the coupling between components in complex instruments prior to integration. 

\begin{acknowledgements}
The authors thank Ronald Hesper for his contributions to the hardware system. K. Davis is currently supported by an NSF Astronomy and Astrophysics Postdoctoral Fellowship under award AST-1801983. This work was in part supported by ERC starting grant ERC-2009-StG Grant 240602 TFPA. The contribution of J.J.A. Baselmans is also supported  by the ERC consolidator grant COG 648135 MOSAIC.
\end{acknowledgements}
%\section*{Conflict of interest}
%The authors declare that they have no conflict of interest.

\bibliographystyle{spphys}       % APS-like style for physics
\bibliography{cpx_FPA} % your references Yourfile.bib

\end{document}